\documentclass[pre,aps,superscriptaddress,twocolumn]{revtex4-1}
\usepackage{graphicx}
\usepackage{dcolumn}
\usepackage{subfigure}
\usepackage{bm}
\usepackage{tabularx}
\usepackage[usenames,dvipsnames]{color}
\usepackage[latin1]{inputenc}
\usepackage{esint}

\usepackage{ulem}

\def\be{\begin{equation}}
\def\ee{\end{equation}}
\def\ba{\begin{array}}
\def\ea{\end{array}}
\def\bea{\begin{eqnarray}}
\def\eea{\end{eqnarray}}

\newcommand\Pos{\mathop{\mathcal{P}}}
\def\Mix64{/home/damien/Mesoplasticity/Mix-Plas-MF/N64}
\def\Green64{/home/damien/Mesoplasticity/Green64}

\begin{document}

\title{From depinning transition to plastic yielding of amorphous media:\\
A soft modes perspective}

\date{\today}

\author{Botond Tyukodi}
\affiliation{PMMH, ESPCI/CNRS-UMR 7636/Univ.~Paris 6 UPMC/Univ.~Paris 7 Diderot,\\
10 rue Vauquelin, 75231 Paris cedex 05, France}
\affiliation{
Physics department, University Babe{\c{s}}-Bolyai, Cluj, Romania}
\author{Sylvain Patinet}
\affiliation{PMMH, ESPCI/CNRS-UMR 7636/Univ.~Paris 6 UPMC/Univ.~Paris 7 Diderot,\\
10 rue Vauquelin, 75231 Paris cedex 05, France}
\author{St\'ephane Roux}
\affiliation{
LMT, ENS-Cachan/CNRS-UMR 8535/Univ.~Paris-Saclay,\\
61 Avenue du Pr\'esident Wilson, 94235 Cachan cedex, France}
\author{Damien Vandembroucq}
\affiliation{PMMH, ESPCI/CNRS-UMR 7636/Univ.~Paris 6 UPMC/Univ.~Paris 7 Diderot,\\
10 rue Vauquelin, 75231 Paris cedex 05, France}
\begin{abstract}
A mesoscopic model of amorphous plasticity is discussed in the
context of depinning models. After embedding in a $d+1$ dimensional
space, where the accumulated plastic strain lives along the additional
dimension, the gradual plastic deformation of amorphous media can be regarded as the motion of an elastic manifold in a disordered landscape. While the
associated depinning transition  leads to scaling properties, the
quadrupolar Eshelby interactions at play in amorphous plasticity
induce specific additional features like shear-banding and weak
ergodicity break-down. The latters are shown to be controlled
by the existence of soft modes of the elastic interaction,
the consequence of which is discussed in the context of depinning.
\end{abstract}

\maketitle

\section{Introduction}
Most liquids flow as soon as they experience shear stress.
In contrast many complex fluids (pastes, foams, colloidal suspensions,
etc.) do not flow for shear stresses lower than some threshold yield limit.
The rheological behavior of these yield-stress fluids parallels the
plasticity of amorphous solids (oxide and metallic glasses, polymers,
etc.). Both families of materials exhibit a rich phenomenology. Close
to the yielding threshold, critical-like behaviors are observed:
avalanches~\cite{Wang-PRL10,Antonaglia-PRL14}, growth of a correlation
length scale~\cite{Goyon-Nat08}, Hershell-Bulkley
law~\cite{Coussot-book14}... In parallel other properties are
reminiscent of glassy phenomena: e.g. thermal~\cite{Cheng-ActaMat08}
and
mechanical~\cite{PMMCVB-JACS06,*VDCPBCM-JPCM08,Egami-AEM08,Revesz-APL08}
history dependence. In the same spirit, strain
localization~\cite{Lewandowski-NatMat06,Divoux-SM11}, a phenomenon of
crucial technological interest (since it controls the mechanical
strength) can be analyzed as an ergodicity break-down process: plastic
activity is trapped in a very limited sub-region of the phase
space~\cite{Torok-PRL00}.

These two phenomenological archetypes (criticality and glass
transition) have motivated parallel modeling efforts. Building on
trap models~\cite{Bouchaud-JPI92} designed to capture ergodicity
breaking and aging at glass transition, Sollich et
al.~\cite{Sollich-PRL97,*Sollich-PRE98,Fielding-SM09,*Fielding-PRL11}
developed Soft Glassy Rheology (SGR) models and could associate
different rheological behaviors of complex fluids to a parameter of
their model, an effective temperature associated to mechanical noise
(see a recent discussion in \cite{Nicolas-EPL14}).
{A different glassy approach has been pursued by Bouchbinder and
  Langer\cite{Bouchbinder-PRE09a,*Bouchbinder-PRE09b,*Bouchbinder-PRE09c}
  who extended the Shear-Transformation-Zone
  theory\cite{FalkLanger-PRE98} to explicitly account for an effective
  temperature related to the slow configurational degrees of freedom
  of the glassy material under shear.  }

The need to go beyond mean field description and understand the
crucial effect of elastic interactions associated to the localized
rearrangements (Eshelby
events)~\cite{Spaepen-ActaMet77,Argon-ActaMet79,FalkLanger-PRE98,Maloney-PRL04b,Tanguy-EPJE06}
responsible of amorphous plasticity has early led to the development
of mesoscopic models accounting for these
interactions~\cite{BulatovArgon94a,*BulatovArgon94b,*BulatovArgon94c}. This
effort of modeling amorphous plasticity and/or rheology of complex
fluids at mesoscopic scale has, since then, been very active
\cite{BVR-PRL02,TPVR-PRE11,*TPVR-Meso12,Picard-PRE02,*Picard-PRE05,Lemaitre-preprint06,Jagla-PRE07,Dahmen-PRL09,Homer-PRB10,Martens-PRL11,Budrikis-PRE13,Homer-ActaMat14,Nicolas-SM14,Wyart-EPL14,Wyart-PNAS14}.
As early noticed in~\cite{BVR-PRL02}, the competition at play in
mesoscopic models between microscopic disorder and elastic interaction
strongly reminds the physics of the depinning
transition~\cite{Fisher-PR98,*Kardar-PR98} that naturally entails
critical features. Recently summarized in Ref.~\cite{Wyart-PNAS14},
most features of the associated scaling phenomenology predicted by
depinning-like models of amorphous plasticity have been observed
numerically~\cite{Maloney-PRL04a,Argon-PRB05b,Maloney-JPCM08,*Maloney-PRL09,Salerno-PRL12,*Salerno-PRE13}
and experimentally~\cite{Wang-PRL10,Antonaglia-PRL14}.

{Noteworthily, some of the key non-ergodic features (e.g. aging and
shear-banding\cite{VR-PRB11,*BV-AIP13,Martens-SM12,Homer-ActaMat14}) can be also
recovered within the framework of the mesoscopic elastoplastic models.}
This has raised the question of the precise link
with the depinning transition. In particular, the crucial effect of
the non-positiveness of the quadrupolar elastic interaction induced by
individual plastic events has been questioned. 
Recently Lin {\it et al.}~\cite{Wyart-PNAS14} have shown the necessity
of three independent exponents (instead of two for standard depinning)
to account for the scaling properties of mesoscopic models of
amorphous plasticity.

Here we show that the specific features observed in elasto-plastic
models are controlled by the presence of multiple soft modes of the quadrupolar
elastic interaction. 
{Note that the presence of such soft modes is {\em not} an artifact of
  lattice discretization or of a specific numerical
  implementation~\cite{Budrikis-PRE13}. In the present perspective,
  shear bands directly result of the Eshelby interaction symmetry
  i.e. extended modes of plastic deformation that satisfy
  compatibility and consequently induce no internal stress.} This
property, absent in classical depinning models, has dramatic effects
on the stability, the dependence on initial conditions as well as the
ergodicity properties of plastic yielding models.

In the following we present in section I the details of the mesoscopic
models of amorphous plasticity. We give a particular emphasis on the
comparison with the models of depinning an elastic manifold in a
random landscape. The emergence of anisotropic elastic interactions
associated to local plastic inclusions is discussed. In section III, a
comparison is presented between numerical results on strain
fluctuations obtained with Mean-Field (MF) and ``Eshelby'' anisotropic
elastic kernels\cite{Eshelby57,Picard-EPJE04}. In section IV, a
Fourier space analysis allows us to unveil the presence of multiple
soft modes of the Eshelby elastic interactions. We show in section V
that this soft mode analysis sheds a new light on the diffusion and
shear-banding behaviors of the mesoscopic models of amorphous
plasticity. Our main results are finally summarized in section VI.


\section{Depinning-like models for amorphous plasticity} 

\subsection{A scalar mesoscopic model}

Here we restrict ourselves to a simple scalar
case~\cite{TPVR-Meso12}. Assuming, bi-axial loading conditions, we
define respectively for stress and strain the scalar quantities
$\sigma= \sigma_{yy}- \sigma_{xx}$, $\varepsilon= \varepsilon_{yy}-
\varepsilon_{xx}$ from their tensor counterparts. The material is
discretized on lattice at a mesoscopic scale $\ell$ and is assumed to
be elastically homogeneous. A simple plastic criterion is defined from
the comparison between the local values of the scalar equivalent
stress field $\sigma(\mathbf{r},\{\varepsilon^{pl}\})=\sigma^{ext}+
\sigma^{int}[\mathbf{r},\{\varepsilon^{pl}\}]$ with a threshold stress
$\sigma^c[\mathbf{r},\varepsilon^{pl}(\mathbf{r})]$. The local stress
$\sigma$ results from the addition of a spatially uniform external
stress $\sigma^{ext}$ and of a spatially fluctuating internal stress
$\sigma^{int}$ due to the successive plastic rearrangements mediated
by the elastic interactions. Here the local stress threshold
$\sigma^c$ encodes the disordered nature of the structure, it depends
both on space and on the local value of the plastic strain
$\varepsilon^{pl}$.

From this local criterion a simple equation can be written to
model the evolution of the plastic strain field:
\be
\partial_t \varepsilon^{pl}(\mathbf{r},t) = \Pos\left(
\sigma^{ext}\!+G^{el}\!*\!\varepsilon^{pl}(\mathbf{r},t)
\!-\!\sigma^{c}[(\mathbf{r},\varepsilon^{pl}(\mathbf{r},t)]\right)
\label{eq-mesoplast}
\ee
Here the threshold dynamics is accounted for by the positive part
function $\Pos(\cdot)$ such that $\Pos(x)=x$ if $x>0$ and $\Pos(x)=0$
if not.

The heterogeneity of the plastic yield stress at mesocopic scale is
represented by the quenched variable $\sigma^{c}$. The latter is
defined by its average $\langle\sigma^{c}\rangle=\overline{\sigma^c}$
and its correlations
$\langle\sigma^{c}(\mathbf{r},z)\sigma^{c}(\mathbf{r}+\mathbf{\delta
  r},\varepsilon^{pl}+\delta \varepsilon^{pl})\rangle=\varsigma^2f(\mathbf{\delta r})g(\delta \varepsilon^{pl})$
where $\varsigma^2$ gives the variance.  Short-range correlations are
considered, namely, $f(\mathbf{\delta r}) \to 0$ if $|\mathbf{\delta
  r}|\gg\ell$ and $g(\delta z)\to 0$ if $|\delta \varepsilon^{pl}|\gg e_0$. The length
scale $\ell$ is given by the mesoscopic scale at which coarse-graining
is performed. The strain scale $e_0$ corresponds
to the typical plastic strain induced by elementary plastic events.

Finally the internal stress $\sigma^{int}$ is represented through a
convolution of the plastic strain field $\varepsilon^{pl}$ and the
elastic kernel $G^{el}$ associated with the reaction of the matrix to
a unit local plastic strain: $\sigma^{int}=G^{el}*\varepsilon^{pl}$
(Eshelby inclusion~\cite{Eshelby57}). The properties of this
long-ranged and anisotropic elastic interaction are discussed in more
details in sub-section~\ref{elastic-interaction}

Instead of directly integrating Eq.~(\ref{eq-mesoplast}), an extremal
dynamics algorithm of the model discretized on a lattice is
implemented~\cite{BVR-PRL02}. Only one site (the weakest one)
experiences plastic deformation at each iteration step.  The external
stress is adjusted accordingly. Such an algorithm corresponds to
shearing the system at a vanishing strain rate and is very close in
spirit to the athermal quasi-static protocols under conditions of
imposed strain developed in atomistic
simulations\cite{Maloney-PRL04a,Maloney-PRE06}.


\begin{figure}[t]
\begin{center}
\includegraphics[width=0.98\columnwidth]{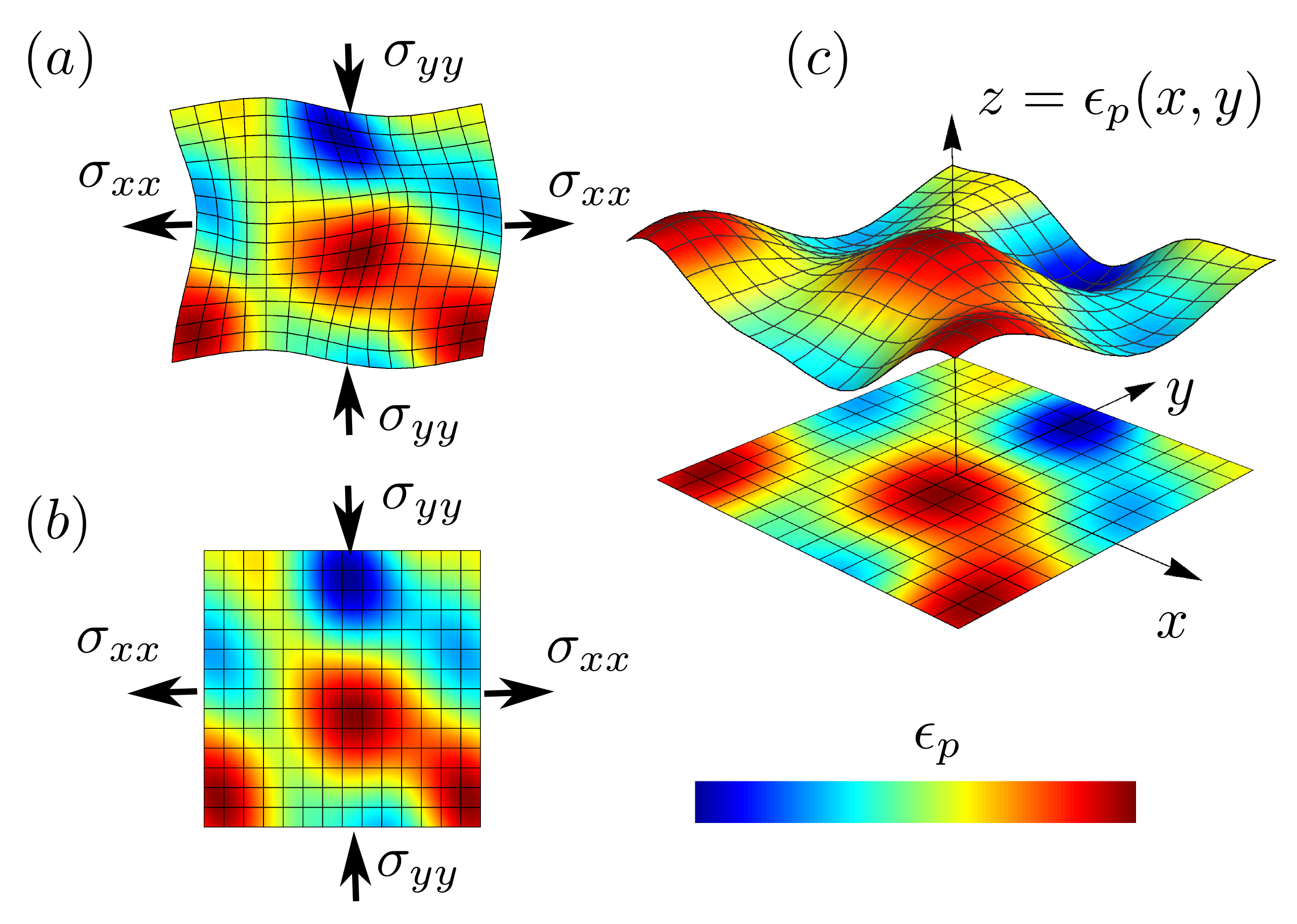}
\end{center}
    \caption{Sketch of a 2d amorphous material upon bi-axial loading. (a) The mesh is deformed according to the displacement.
The associated strain has a reversible elastic contribution and an irreversible plastic contribution.  The latter is represented according to the color scale. (b) The plastic strain field (colors) is represented on the un-deformed reference frame. (c) The plastic strain field can be represented as a $d$-dimensional manifold moving in a $d+1$ space.}
\label{depinning-vs-plastic-yielding}
\end{figure}

\subsection{From plastic yielding to depinning}

In the framework of upscaling amorphous plasticity from the
microscopic to the macroscopic scales~\cite{RTV-MSMSE11}, the equation
of evolution (\ref{eq-mesoplast}) can be understood in one of the two
ways. First it can be seen as presented above i.e. as a description of
the (visco-)plastic dynamics of a plastically heterogeneous material,
discretized at scale $\ell$.

Second, such a threshold dynamics also naturally emerges after
coarse-graining (in the direction of motion) from the equation of
evolution of a driven elastic manifold in a continuous random
landscape. In order to illustrate this direct mapping to depinning we
discuss in the following the geometry of the equivalent manifold and
the emergence of the threshold dynamics associated to the
multistability of the elastic interface.

Let us recall the equation of evolution of the overdamped motion of an
elastic manifold $h({\mathbf x})$ in a random
landscape~\cite{Fisher-PR98,*Kardar-PR98}:
\be
  \partial_t h(\mathbf{r},t) = f^{ext}(t)+G^{el}\!*\!h(\mathbf{r},t)
  -\frac{\partial U}{\partial h}
  \left[\mathbf{r},h(\mathbf{r},t)\right]
\label{eq-depinning}
\ee
Here $f^{ext}$ stands for the external driving force, $G^{el}*h$ for
the elastic restoration force and $U$ is a random potential such that
$\langle \partial_\mathbf{r} U \rangle =0$ and $\langle
\partial_\mathbf{r} U(\mathbf{r},z) \partial_\mathbf{r}
U(\mathbf{r'},z') \rangle =\varsigma^2f(\delta r/\ell)g(\delta z/e_0)$
where $\ell$ and $e_0$ give the correlation lengths along the manifold
and in the direction of propagation, respectively.

The present depinning equation is very close to
Eq. (\ref{eq-mesoplast}) proposed to model amorphous plasticity. In
the latter the external stress plays the role of the driving force for
the depinning, the elastic kernel associated to the Eshelby inclusions
corresponds to the elastic restoration force and the disordered stress
thresholds are associated to the random potential.

To illustrate more clearly the direct analogy between deformation
under shear and motion of an elastic manifold we give here a simple
geometric interpretation. Let us consider the plastic strain field
$\varepsilon_p(\mathbf{r})$ of a $d$-dimensional material.  As
sketched on Fig.~\ref{depinning-vs-plastic-yielding}, we can define an
extra coordinate $z$, orthogonal to the space variable $\mathbf{r}$
after embedding in a $d+1$ dimensional space. The equation
$z=\varepsilon^{pl}(\mathbf{r})$ thus defines an elastic manifold
whose propagation in the random landscape $\sigma^{c}
[\mathbf{r},\varepsilon^{pl}(\mathbf{r},t)]$ is governed by
Eq.~(\ref{eq-mesoplast}).

An obvious difference still remains between the two equations. While
the the depinning equation (\ref{eq-depinning}) models a continuous
evolution, the equation (\ref{eq-mesoplast}) shows a discontinuous
threshold dynamics, here encoded by the presence of the $\Pos(.)$
function. We argue here that, far from being different in nature, such
a threshold dynamics is a direct outcome of the competition between
elasticity and disorder upon coarse-graining in the direction of
propagation..

\begin{figure}
\begin{center}
\includegraphics[width=0.98\columnwidth]{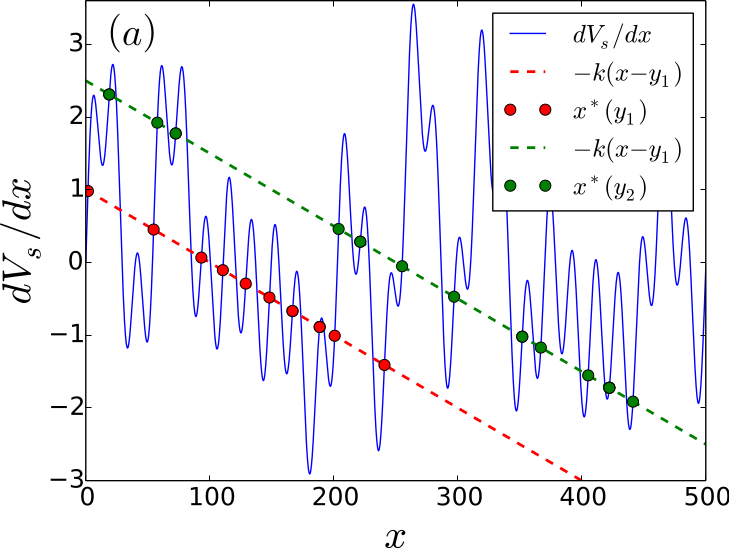}
\includegraphics[width=0.98\columnwidth]{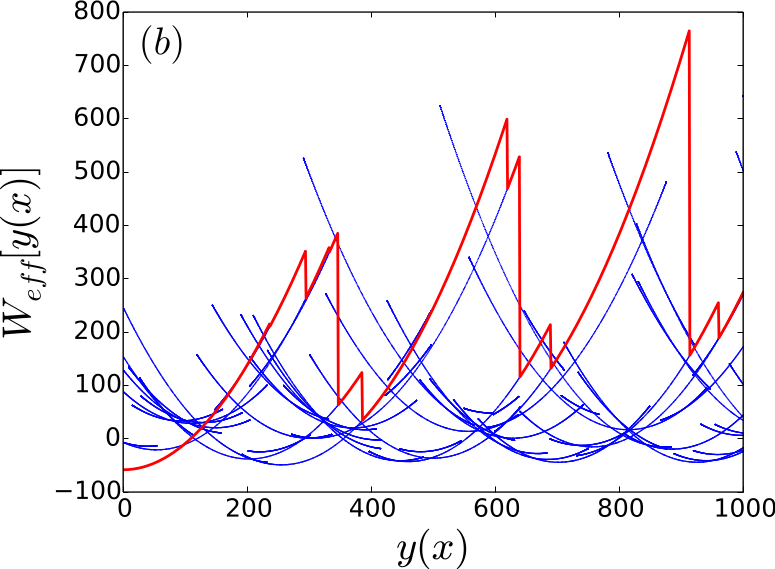}
\includegraphics[width=0.98\columnwidth]{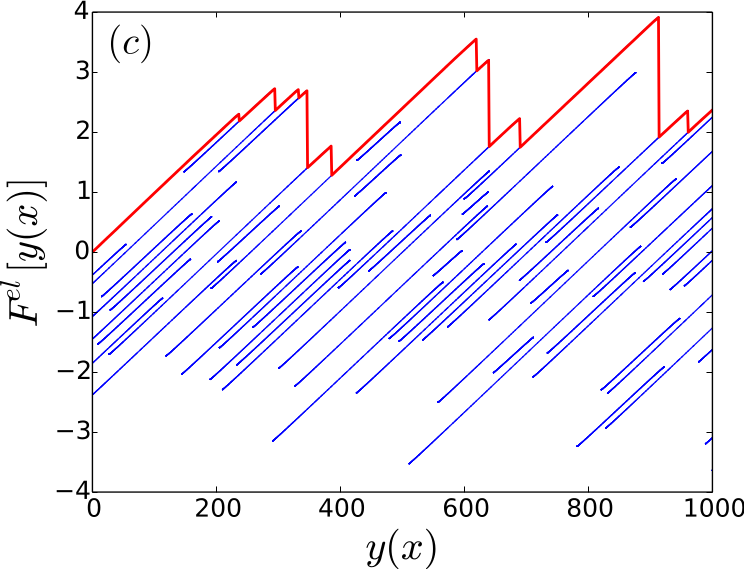}
\end{center}
\caption{Phenomenology of the motion of a particle of position $x$
  driven by a spring of position $y$ in a one-dimensional random
  potential $V(x)$: (a) Graphical representation of the multiplicity
  of the solutions of the equilibrium equation $V'(x) = -k(x-y)$ for
  two positions of the spring. (b) Parametric represntation of the
  complex effective potential $W_{eff}(y)$ seen by the spring and
  representaion (in red) of on eparticular trajectory. (c) Associated
  represntation of the force landscape. Jumps in the potential
  $W_{eff}$ are associated to force thresholds.}
\label{multistability}
\end{figure}

In order to give more support to the latter statement we resort in the
following to a simple example early developed in the close contexts of
solid
friction~\cite{Caroli-EPJB98,*Tanguy-PRE97,*Baumberger-AdvPhys06} and
rate independent plasticity~\cite{Trusk-JMPS05}, the over-damped
dynamics of an isolated point driven into a one-dimensional random
potential:
\bea
{\partial_t}x &= 
-{\partial_x} \left[ \frac{k}{2}\left(x-y\right)^2 \right]
- V'(x)\\
\nonumber &=  -{\partial_x} W(x,y)
\eea

where $V$ is a random potential such that $\langle V' (x) V'
(x')\rangle = \varsigma^2 f[(x-x')/e_0]$ where $f(u) \to 0$ for
$|u| \ll 1$. Here $y$ denotes the external driving (e.g. at finite
velocity $y=vt$) and $k$ is the strength of the confining potential
(the stiffness of the spring driving the system).

Such a system of total energy $W(x,y) = k(x-y)^2/2 +V(x)$ is known to
exhibit multistability when disorder overcomes elasticity. Namely if
$k\varsigma/e_0>1$, for every $y$ position, one and only one position
$x^*(y)$ satisfies equilibrium and stability conditions: $\partial_x
W(x,y)=0$ and $\partial^2_{xx} W(x,y)>0$. An effective potential
$W_{\mathrm{eff}}(y) = W[x^*(y)]$ can then be defined unambiguously.

Conversely, as illustrated in Fig.~\ref{multistability}a that shows
graphical solutions of the equilibrium equation $-k(x-y)=V'(x)$, for
$k\varsigma/e_0<1$ the potential $W$ is characterized by a large
number of local minima and several stable positions $x^*(y)$ of local
equilibrium can be found for a given position $y$.

Still, it is possible in this multistability case to resort to a
parametric representation and to build an effective potential
$W_{\mathrm{eff}} [y^*(x)] = W[x,y^*(x)]$ associated to the multiple
minima. As shown in Fig.~\ref{multistability}b, the stable branches of
this effective potential consist of series of truncated
parabolas. Upon driving, the system jumps from one local minimum $i$
to another one $j$ as soon as a force exceeds the threshold value
$f_i=-V(x^M_i)$ associated to the upper bound $x^M_i$ of the basin of
attraction of the minimum $i$ (the intersection with the next
parabola).  One obviously recovers here the phenomenology of the
instability inducing local rearrangements at the atomic scale in
amorphous plasticity~\cite{Maloney-PRE06}.

An example of such an (history-dependent) trajectory made of series of
micro-instabilities is shown in Fig.~\ref{multistability}b and
\ref{multistability}c. A threshold dynamics thus directly emerges from
this simple case of an isolated defect. In particular, as shown in
Fig.~\ref{multistability}c it is clear that upon coarse-graining at
scale $\xi$, the dynamics of jumps between basins is entirely
controled by the series of threshold forces $f_i$.

The phenomenology remains unchanged when dealing with more complex
objects like manifolds. Rather than the stiffness of an external
device, the disorder has in this case to be compared with the internal
elasticity of the manifold. See e.g. Ref.~\cite{PVR-PRL13} for a
recent discussion in the context of crack front propagation. Note that
the non-regularity of the effective potential induced by
multistability is likely to be related to the emergence of a cusp in
the correlator of depinning forces observed under
renormalization~\cite{Rosso-PRB07}.

The present model of amorphous plasticity this appears to belong to
the wider class of depinning models. We discuss in the next section to
what extent the peculiar nature of the Eshelby elastic interaction
associated with plasticity does affect the phenomenology of
depinning.

\subsection{A peculiar elastic interaction \label{elastic-interaction}}

The occurrence of a plastic local rearrangement in the amorphous
structure inuces internal stresses due to the reaction of the elastic
surrounding matrix. This results in a stress relaxation of the region
that rearranged and in an anisotropic long-ranged stress field in the
outer matrix. This elastic interaction is very peculiar. In
particular, it is non strictly positive: the sign depends on the
direction. The elastic interaction thus either favors or unfavors the
occurence of future plastic events depending on their position.

The exact internal stress field obviously depends on the details of
the rearrangement of the amorphous structure. A classical
approximation consists in resorting to a continuum mechanics analysis
and in using the solution of the stress induced by a plastic inclusion
early proposed by Eshelby~\cite{Eshelby57}. More precisely,
independently on the precise shape of the inclusion, only the dominant
contribution of the internal stress in the far-field is considered.

\begin{figure}
\begin{center}
\includegraphics[height=0.36\columnwidth]{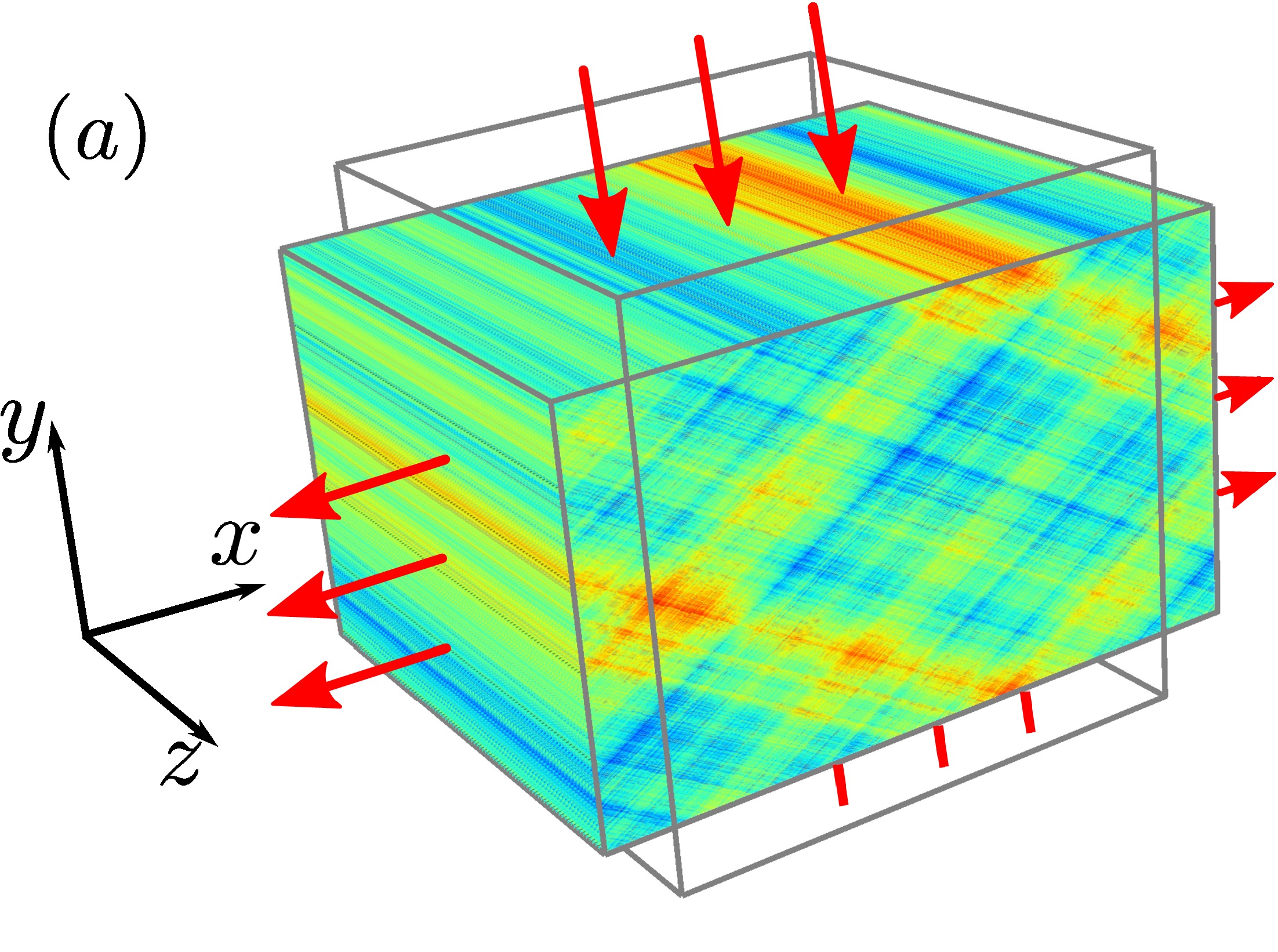}
\includegraphics[height=0.37\columnwidth]{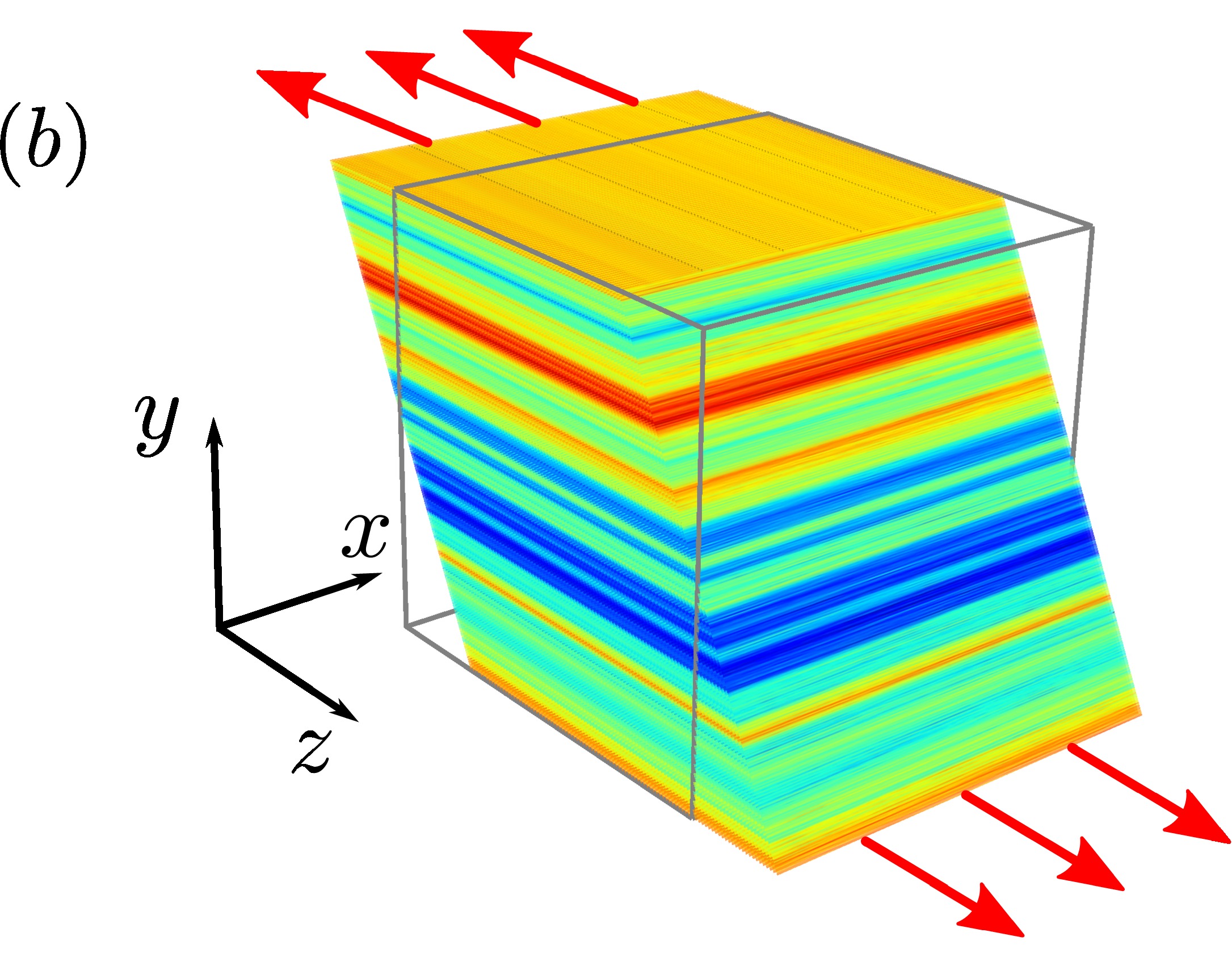}
\end{center}
\caption{Sketch of two kinds of shear geometry: (a) Plane shear
  geometry; (b) Antiplane shear geometry. In both cases the strain
  field is invariant along the $z$-axis. The color scale gives the
  amplitude in the $xy$ plane of the plastic strain fields (a):
  $\varepsilon^{pl}=\varepsilon^{pl}_{yy}-\varepsilon^{pl}_{xx}$ and
  (b) $\varepsilon^{pl}_{yz}$. A quadrupolar symmetry is observed in
  the plane shear case (a) while a dipolar symmetry is observed in
  antiplane shear case (b).}
\label{plane-antiplane}
\end{figure}

In the plane strain geometry considered in
  Fig.~\ref{plane-antiplane}a, a pure shear plastic inclusion induces
  a long-range internal stress characterized by a quadrupolar
  symmetry. In an infinite medium, the dominant term in the far-field
  and the mean stress drop in the inclusion can be written respectively: 
\be
G_Q( \mathbf{r}) = -2\mu^* a^2\varepsilon_p \frac{\cos(4\theta)}{r^2} \;,\quad 
G_Q( \mathbf{0}) =-\mu^* \varepsilon_p \;.
\label{Eshelby-farfield}
\ee where $\mu^*$ is an effective elastic modulus, $a$ the mean radius
of the inclusion and $\varepsilon_p$ the mean plastic strain
experienced by the inclusion.  Here the subscript {\it Q} refers to
the quadrupolar symmetry. Note that the amplitude of this quadrupolar
elastic interaction is controlled by the product of the ``volume'' of the
inclusion by the mean plastic strain.

For the numerical implementation, bi-periodic boundary conditions
  are considered and following Ref.~\cite{TPVR-Meso12}, a quadrupolar
  lattice Green function is defined from the following expression in
  the Fourier space: 
\be
   \widetilde{G_Q}(p,q) =  -A\left[\cos (4\theta_{pq}) +1\right]\;, \quad
\widetilde{G_Q}(0,0) = 0 \;.
\label{Eshelby-Fourier}
\ee
where $\theta_{pq}$ is the polar angle and $(p,q)$ the wavevector in
Fourier space. While the first term directly stems from the
quadrupolar symmetry of the Eshelby far field expression
(\ref{Eshelby-farfield}), the null value of the zero frequency term
$\widetilde{G_Q}(0,0)$ is required by a stationarity condition : a
spatially uniform plastic strain induces no internal stress. In other
words, no plastic incompatibilities are generated because of the
assumption of uniform elastic moduli. When translated to discrete
form, it means that $\sum_{i,j} G_Q(i,j)=0$, henceforth this condition
directly imposes the value of the latice Green function at the origin,
i.e. the stress drop:
\bea
 G_Q(0,0) &= &- \sum_{(i,j)\ne (0,0)} G_Q(i,j) \\
\nonumber
&= &-\frac{A}{N^2} \sum_{(p,q)\ne (0,0)}\left[\cos (4\theta_{pq})+1\right]
\eea 
The prefactor $A$ has the dimension of an elastic modulus. Here it is
chosen so that the local stress relaxation in the site that
experienced a unit plastic deformation is unity: $G_{Q}(0,0)=-1$.

In the plane shear strain geometry (invariant along the
$z$-coordinate) illustrated in Fig.~\ref{plane-antiplane}a, the
quadrupolar elastic interaction $G_{Q}$ is positive in the directions
at $\pm \pi/4$ and negative in the directions at $0$ and $\pi/2$. The
associated plastic strain field is thus orientated along the diagonals
of the $x,y$ plane.

For the sake of completeness, we also illustrate in
Fig.~\ref{plane-antiplane}b another loading geometry: the antiplane
shear geometry. Here the strain field is again invariant along the
$z$-axis but the system is sheared along the $yz$ direction so that
only $u_z$ the $z$-component of the displacement field is non zero and
the strain component of interest is $\varepsilon_{yz}(x,y)=\partial_y
u_z(x,y)$. Within this antiplane geometry early studied in
Ref.~\cite{BVR-PRL02}, the elastic interaction associated to a plastic
inclusion obeys a dipolar geometry: $G_D( \mathbf{r}) = A
\cos(2\theta)/r^2$ so that the plastic strain field is orientated
along the $x$ direction. The specificity of this loading geometry will
be further discussed in section~\ref{plane-vs-antiplane}.

Due to their long range character, it may be tempting to approximate
the ``Eshelby'' elastic interaction by a simple Mean-Field (MF)
interaction~\cite{Dahmen-PRL09}: $G_{MF}(\bm r_{ij})=1/(N^2-1)$ if
$|\bm r|\neq 0$ and $G_{MF}(\bm 0)=-1$. The latter will be used (all
other parameters being kept constant) to illustrate the expected
behavior of a standard reference depinning model. In the following, we
compare the respective effects of Mean-Field and quadrupolar
interactions on some specific properties of amorphous plasticity,
i.e. strain diffusion and localization.  In order to to investigate
the origin of the specific effects of the ``Eshelby'' elastic kernel,
we also define a weighted average of two propagators: $G_a= (1-a)
G_{Q} + a G_{MF}$ where the parameter $a$ gives the relative weight of
the mean field.  For moderate values of $a$, the quadrupolar symmetry
is mainly preserved in the sense that the Green function remains
strictly negative in the $0$ and $\pi/2$ directions.

\section{Mean-Field Depinning vs plasticity}

\begin{figure}[t]
\begin{center}
\includegraphics[width=0.95\columnwidth]{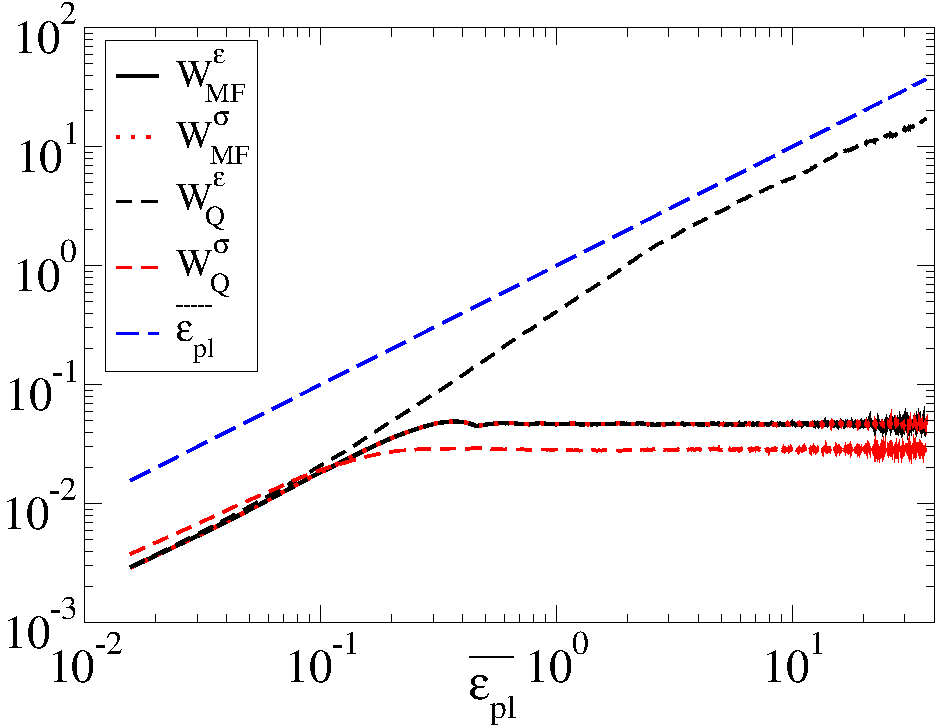}
\end{center}
\caption{Plastic train variance $W^\varepsilon$ and elastic stress
  variance $W^\sigma$ vs cumulated plastic strain $\overline{
    \varepsilon^{pl} }$ for a quadrupolar propagator $G_Q$ and a
  Mean-Field $G_{MF}$. A linear behavior is represented for
  comparison.}
\label{diffusion1}
\end{figure}
\begin{figure}
\begin{center}
\includegraphics[width=0.95\columnwidth]{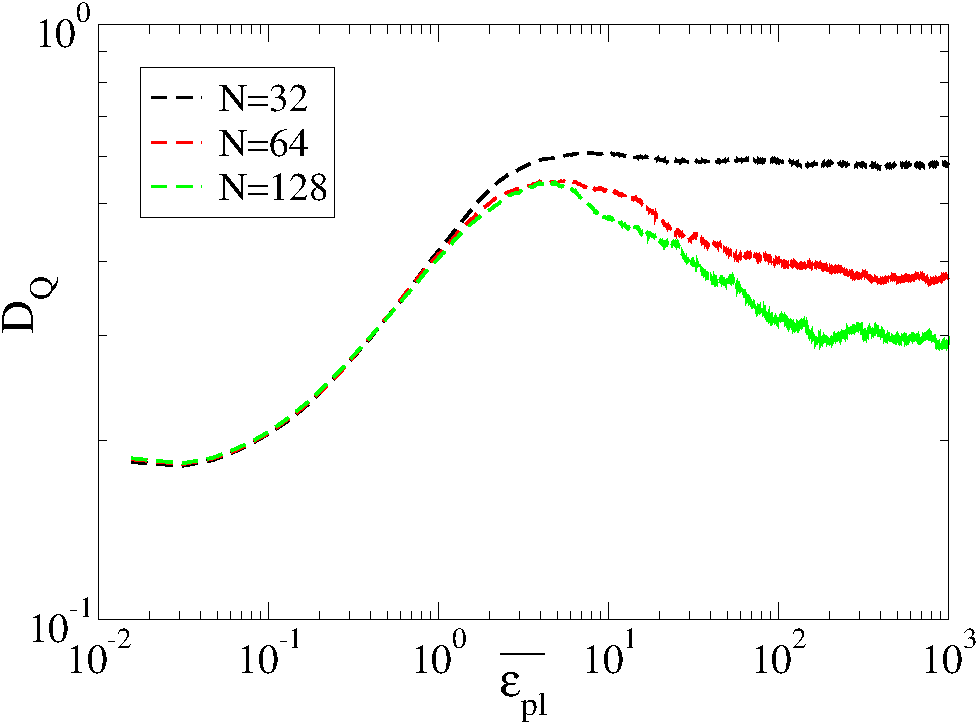}
\end{center}
\caption{Size-dependent behavior of the plastic strain diffusivity
  $D_Q=W_Q^\varepsilon/\overline{\varepsilon^{pl}}$ obtained with the
  quadrupolar kernel $G_Q$ for sizes $N=32,\;64,\;128$ with
  $M=1000,\;100,\;30$ realizations respectively. The larger the
  system, the longer the anomalous sub-diffusive behavior.}
\label{diffusivity}
\end{figure}

\subsection{Family-Vicsek scaling vs Diffusion}

We first discuss the behavior of the variance of the plastic strain
$W^\varepsilon=\langle \overline{|\delta \varepsilon^{pl}|^2} \rangle$
where we defined the spatial fluctuation of the plastic strain field
$\delta \varepsilon^{pl} =\varepsilon^{pl} -
\overline{\varepsilon^{pl}}$). Here $\overline{X}$ and $\langle X
\rangle$ denote the spatial average and the ensemble average of the
variable $X$, respectively. We show in Fig.~\ref{diffusion1}~(a) the
evolution of the variance $W^\varepsilon$ with respect to the mean
plastic strain $\overline{\varepsilon^{pl}}$. In the context of
depinning, as illustrated in Fig~\ref{depinning-vs-plastic-yielding},
$W^\varepsilon$ is nothing but the width of the propagating
interface. Moreover, in the framework of extremal dynamics used here,
the mean plastic strain $\overline{\varepsilon^{pl}}$ defines a
fictive time directly associated to the total number of iterations. It
is thus legitimate to discuss our results in the framework of the
classical Family-Vicsek
scaling~\cite{Narayan-PRB93,Tanguy-PRE98,Narayan-PRE00} for interface
growth. The latter predicts first for the interface width $W$, a power
law growth $W\propto t^\alpha$ up to a time scale $\tau\propto L^z$
such that the correlation length $\xi$ has reached the system size
$\xi(\tau)\approx L$ and beyond which saturation is obtained.

\begin{figure}[t]
\begin{center}
\includegraphics[width=0.95\columnwidth]{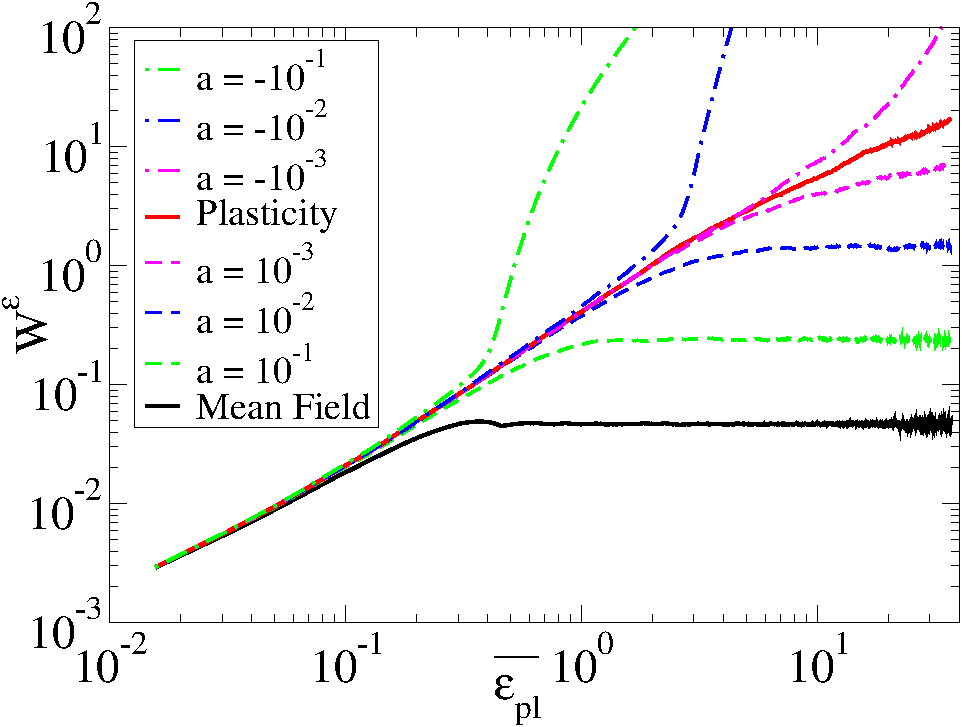}
\end{center}
\caption{Strain variance (equivalently interface width) vs
  cumulated plastic strain $\overline{\varepsilon_{pl}}$ for 8
  different propagators: $G_Q$, $G_a$ ($a=\pm 10^{-1}, \pm 10^{-2},
  \pm 10^{-3}$), $G_{MF}$.}
\label{width-Mix}
\end{figure}

\begin{figure*}
\begin{center}
 \includegraphics[height=0.63\columnwidth]{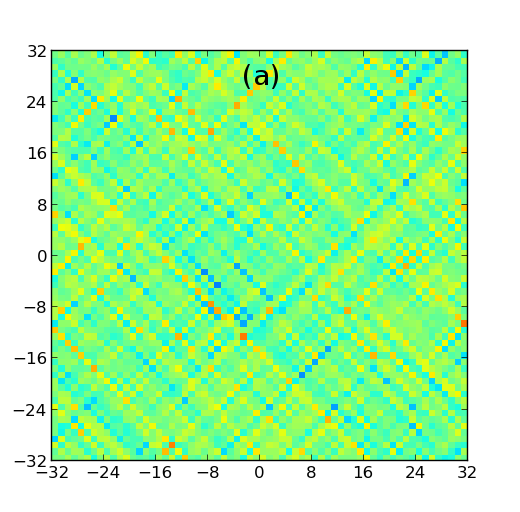}
 \hspace{-10pt}
 \includegraphics[height=0.63\columnwidth]{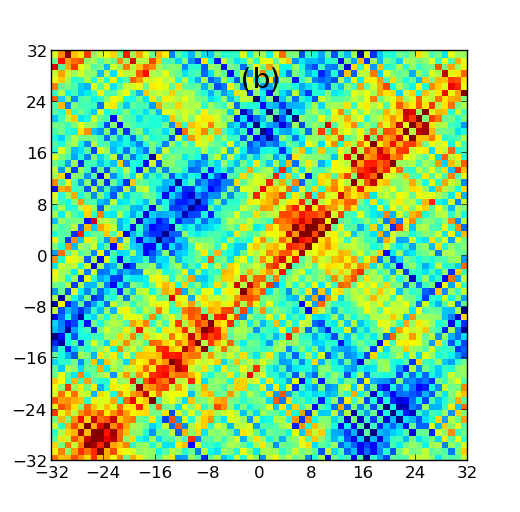}
 \hspace{-10pt}
 \includegraphics[height=0.63\columnwidth]{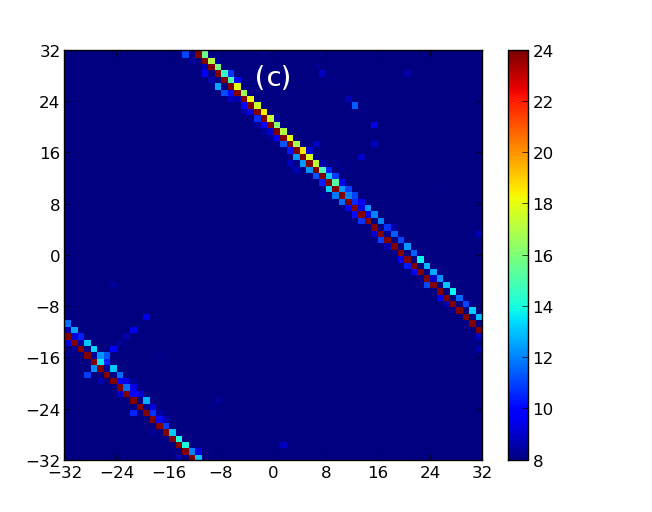}
\end{center}
\caption{Maps of plastic strain field
  obtained for a mere quadrupolar elastic interaction (b), and with a
  positive (a) and a negative (c) MF contribution $a=\pm 10^{-2}$ for
  $\overline{\varepsilon_{pl}}\approx 5$, past the transient
  regime.  The same color scale has been used in the three cases. }
\label{maps-mix}
\end{figure*}

Our numerical results are shown in Fig.~\ref{diffusion1} for
Mean-Field and quadrupolar elastic interactions. As expected, the
classical Family-Vicsek scaling is recovered for the width
$W_{MF}^{\varepsilon}$ obtained in the case of the Mean-Field
depinning.  In the amorphous plasticity case, the first power-law
growth regime is recovered but, past $\xi\approx L$, the interface
width $W_{Q}^{\varepsilon}$ shows no saturation but rather a diffusive
trend~\cite{TPVR-Meso12}. The evolution of the variance $W^{\sigma}$ of
the elastic stress field $\sigma^{el}$ is also shown in the two
cases. Here saturation is recovered in plasticity as well as in MF
depinning. Note that the elastic stress field can be directly obtained
from the plastic strain field from a simple convolution with the
propagator: $\sigma^{el} =G*\varepsilon^{pl}$. The fact that the
diffusive trend at play with the strain field does not show in the
stress fluctuations is a first indication that strain fluctuations are
controlled by soft modes of the elastic interaction.

In order to characterize in more details the diffusive-like behavior
of the plastic strain field obtained with the quadrupolar elastic
interaction, we show in Fig.~\ref{diffusion1}(b) the evolution of the
associated effective difusivity
$D_Q=W_Q^{\varepsilon}/\overline{\varepsilon_{pl}}$. 
This ratio is expected to be
constant for standard diffusion. At very short times, a plateau is
observed; In this very early regime, plastic activity is not
correlated yet. Then the diffusivity $D_Q$ shows a power-law
growth. This simply derives from the fact that in this regime the
growth exponent $\alpha$ is larger than unity: 
$D_Q =W_Q^{\varepsilon}/\overline{\varepsilon_{pl}} \propto |\overline{\varepsilon_{pl}}|^{\alpha-1}$.

The evolution of the diffusivity then shows a strong
size-dependence. For small system size, a simple plateau is obtained,
the diffusivity saturates to a constant value. However for larger
system sizes a long decreasing transient is observed before a
stationary value is obtained. The larger the system, the longer the
transient sub-diffusive regime.

\begin{figure*}[t]
\begin{center}
\includegraphics[width=0.95\columnwidth]{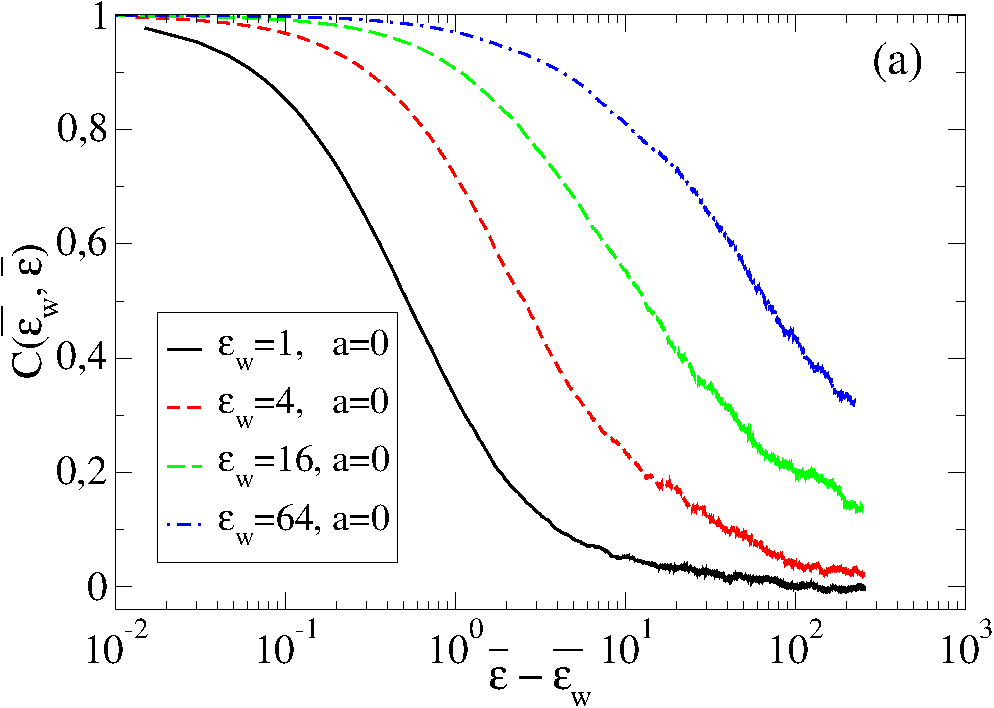}
\includegraphics[width=0.95\columnwidth]{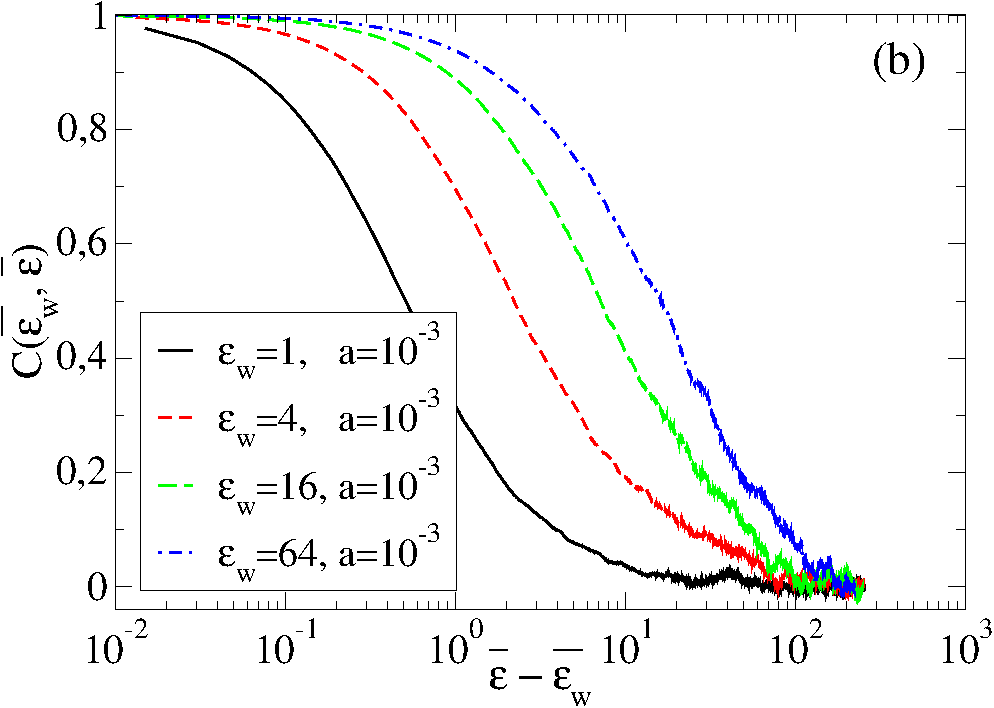}
\includegraphics[width=0.95\columnwidth]{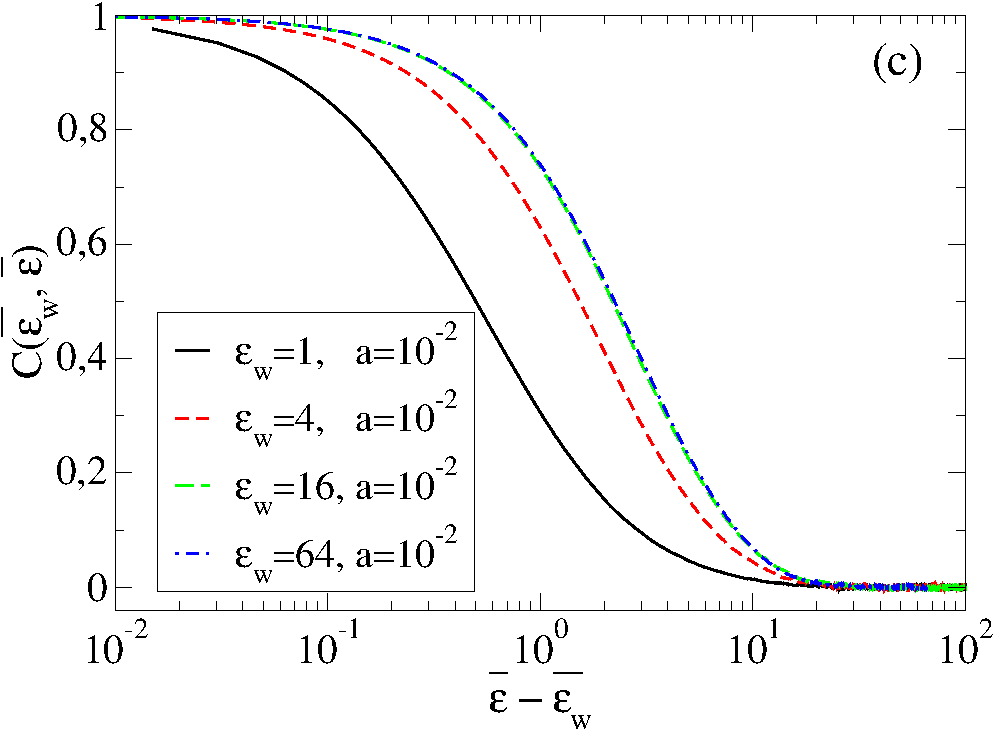}
\includegraphics[width=0.95\columnwidth]{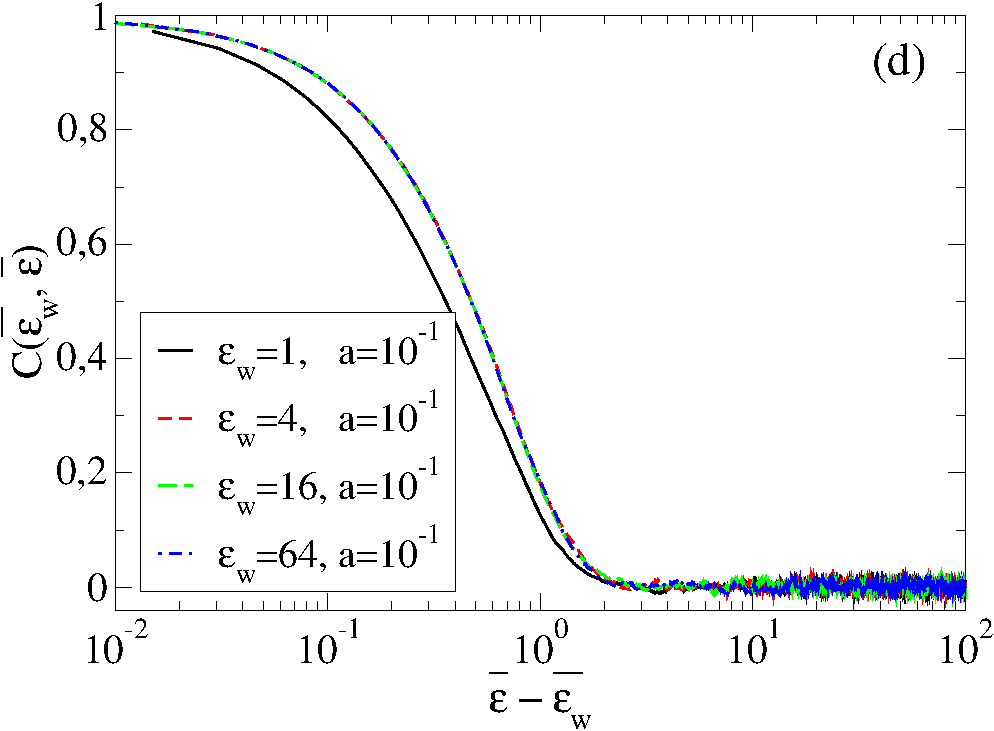}
\end{center}
\caption{ Two-point correlation $C(\overline{\varepsilon_w},
  \overline{\varepsilon})$ of the plastic strain field as a function
  of the cumulated mean plastic strain $\overline{\varepsilon}$ for 4
  ``waiting times'' $\overline{\varepsilon_w} = 1,\;4,\;16,\;64$ and
  for 4 different propagators: $G_Q$ (a), $G_a$ with a Mean-Field
  weight $a=10^{-3}$ (b), $a=10^{-2}$ (c) and $a=10^{-3}$ (d). A clear
  aging effect shows for the quadrupolar propagator $G_Q$: the longer
  the waiting time, the slower the decorrelation. The small MF
  contributions in propagators $G_a$ gradually kills the aging
  behavior.  }
\label{aging}
\end{figure*}

\subsection{Shear-banding and plastic aging}

The nature of the elastic interaction thus strongly affects the
evolution of the spatial fluctuations of the plastic strain field and
in particular the existence of a diffusive regime. In order to get
more insight on the respective effects of the Mean-Field and the
quadrupolar kernels, we now show results obtained with the mixed
kernel $G_a= (1-a) G_{Q} + a G_{MF}$.

In Fig.~\ref{maps-mix}~(top) the  evolution of the interface width
is shown for different (small) values of $a$. It turns out that even the
lowest positive MF contribution is enough to recover saturation at long
times. A transient diffusive regime appears when $a$ tends to zero,
and the level of the final plateau increases accordingly. But when the
interface gets too distorted, if $a>0$ the (low) MF restoring force
eventually stops the divergence of the strain fluctuations. 

A negative MF contribution $a<0$ has the opposite effect: after a
transient diffusive regime, the plastic strain becomes unstable and
its variance diverges very fast. The diffusive regime thus appears to
be a specific feature of the quadrupolar kernel. It lives on the verge
of stability and any mean-field contribution to the elastic kernel
sends the system either toward saturation or ballistic evolution
depending on the sign of $a$.

The strong effect of the MF contribution is also manifest in the
spatial distribution of the plastic strain field. In
Fig.~\ref{maps-mix}~(bottom) maps of the plastic strain are shown for
a cumulated plastic strain $\langle \varepsilon^{pl}\rangle \approx 5$
for $a=-0.01,\;0,\;0.01$ using the same color scale. The plastic case
($a=0.0$) shows a superposition of patterns localized at $\pm \pi/4$
following the symmetry of the quadrupolar kernel. Similar patterns
survive with a positive MF contribution ($a=0.01$) but get very
attenuated (the interface width is much lower). A negative MF
contribution induces conversely a strong localization behavior:
plastic activity is restricted along a unique very thin shear band.

As above mentioned, shear-banding can be analyzed as a kind of
ergodicity breakdown: plastic deformation only visits a sub-part of
the phase space~\cite{Torok-PRL00}. It is thus tempting to analyze the
present model of plastic yielding along these lines.  In
Fig.~\ref{aging} we show two-point correlation functions computed
after various ``waiting times'' $\overline{\varepsilon_w}$ (here the
cumulated plastic strains):
\begin{equation}
C(\overline{\varepsilon_w},\overline{\varepsilon})=
\frac{\langle \overline{\varepsilon^{pl}(x,y,\overline{\varepsilon}) \varepsilon^{pl}(x,y,\overline{\varepsilon_w})} \rangle}{\langle \overline{\varepsilon^2}\rangle^{1/2}\langle \overline{\varepsilon_w^2}  \rangle^{1/2}}
\end{equation}

 For the bare plasticity model, a striking mechanical history effect
 is observed: the larger the waiting time, the larger the
 decorrelation time. Again, the addition of a very small MF
 contribution is enough to destroy this mechanical history dependence.
 Such results are reminiscent of recent studies of depinning
 lines~\cite{Cugliandolo-PRB09} that revealed aging properties but
 only in the roughness growing stage.  Here the saturation of the
 interface roughness is postponed at infinity and aging can persist
 forever. This regime is thus naturally associated to the divergence
 of the interface width.

Note that such an aging behavior may also be observed in a simple
diffusion process.  The diffusion regime at play in amorphous
plasticity is however highly non
trivial~\cite{Maloney-PRL09,TPVR-Meso12}. In particular, as shown in
Fig.~\ref{diffusion1}~(b), for large systems, a very long
sub-diffusive transient regime is obtained i.e. we get
$W^{\varepsilon}_Q \propto \overline{ \varepsilon_{pl}}^\beta$ with
$\beta<1$ over a wide range of strain. This observation again supports
weak ergodicity breakdown. The latter behavior is indeed associated to
sub-diffusion~\cite{Barkai-PRL07c}.


\section{Fourier space and soft modes of the elastic interaction}

The introduction of yet a tiny MF component has thus dramatic
consequences on the localization behavior, a key feature of amorphous
media plasticity. In the following, a rewriting in Fourier space
allows one to emphasize the crucial role of the soft modes of the
propagator in this phenomenon and their connection to plastic
shear-bands.

\subsection{A Fourier representation of depinning}

In the model presented above, the ``Eshelby'' quadrupolar interaction
was defined through its Fourier transform in order to handle periodic
boundary conditions~\cite{TPVR-Meso12}:
\begin{equation}
    \widetilde{G^Q_{pq}} =  A\left(-\cos (4\theta_{pq}) -1\right)
    = - 2A\left( \frac{p^2-q^2}{p^2+q^2}\right)^2 \;,
\end{equation}
where $\theta_{pq}$ is the polar angle and $(p,q)$ the wavevector in Fourier space. $A$ is a constant chosen so that $G(0,0)=-1$.  The Fourier transform of the plastic strain field 
is defined as:
\begin{equation}
    \varepsilon^{pl}_{mn} = \frac{1}{N^2}\sum_{p=-N/2}^{N/2-1}\sum_{q=-N/2}^{N/2-1}
    \widetilde{\varepsilon^{pl}_{pq}}\; e^{-i\frac{2\pi m p }{N}} e^{-i\frac{2\pi nq }{N}} \;.
\end{equation}
The Fourier components of the quadrupolar elastic interaction is thus:
\begin{equation}
\widetilde{\sigma^{el}_{pq}}\!= \widetilde{G^Q_{pq}} \widetilde{\varepsilon^{pl}}_{pq}\!
=\!-2 A \left( \frac{p^2\!-\!q^2}{p^2\!+\!q^2}\right)^2 \widetilde{\varepsilon^{pl}}_{pq}
\;.
\end{equation}
Denoting ${\mathbf e}_{pq} = e^{-i\frac{2\pi m p }{N}}e^{-i\frac{2\pi
    nq }{N}}$ the $(p,q)$ Fourier mode, we get $G^Q*{\mathbf e}_{pq}
= \lambda_{pq} {\mathbf e}_{pq}$ with $\lambda_{pq} =-2 A
[(p^2-q^2)/(p^2+q^2)]^2 $.
%
%
In other terms, the eigenmodes of the Green operator are precisely the Fourier modes, and the associated eigenvalues are the above written $\lambda_{pq}$. This property stems from the translation invariance of the elastic propagator. 

The same property  also holds for the MF propagator: 
\begin{eqnarray}
\nonumber
G^{MF}_{mn} &= &-\delta_m\delta_n +\left( 1-\delta_m\delta_n\right)/(N^2-1)\\
\widetilde{G^{MF}_{pq}} &=&-\frac{N^2}{N^2-1}(1-\delta_p\delta_q)
\label{MF}
\end{eqnarray}
 where $N$ is the linear size of the system.

Let us now discuss the eigenvalue spectrum of the quadrupolar
interaction. One first recognizes the translation mode of zero
eigenvalue $\lambda_{00}=0$. In the classical depinning case (say MF,
Laplacian or power-law in distance) this mode is the only one
characterized by a zero eigenvalue.  It is the signature of the
invariance of the model with respect to a uniform translation of the
manifold along its propagation direction.

\begin{figure}
 \includegraphics[width=0.95\columnwidth]{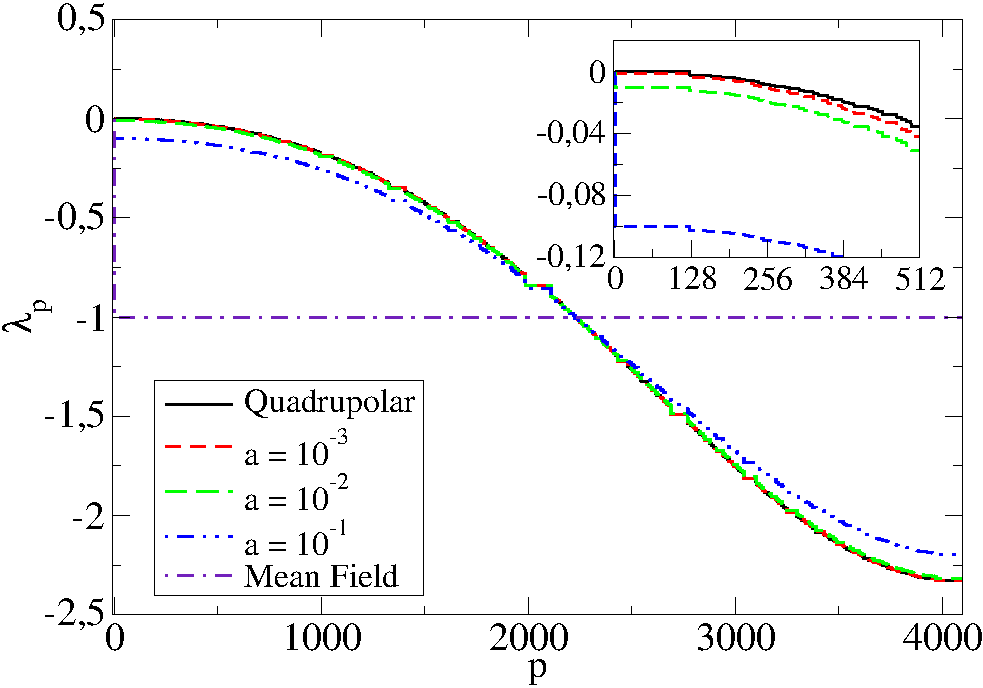}
\caption{Spectrum of eigenvalues of elastic propagators: Mean-Field
  (MF), quadrupolar interaction and MF-weighted quadrupolar
  interactions. Eigenvalues are here simply ranked in the decreasing
  order. The introduction of a fraction $a$ of MF opens a gap between
  the translational mode having a null eigenvalue and the other modes
  $\lambda <0$. The evolution of the gap is zoomed in the inset.}
\label{spectrum}
\end{figure}

In the quadrupolar case, a set of non-trivial eigenmodes are also
characterized by a null eigenvalue. Namely ${\mathbf e}_{p,p}=
e^{-i\frac{2\pi p(m+n) }{N}}$ and ${\mathbf e}_{p,-p}= e^{-i\frac{2\pi
    p(m-n) }{N}}$ with $p \in [-N/2,N/2-1]\backslash \{0\}$. Thus
there is one trivial zero translation eigenmode and $2(N-1)$
non-trivial ones. 

Let us rewrite the the plastic strain field in the Fourier basis using the more condensed form:
\begin{equation}
    \varepsilon^{pl} = \sum_{p,q} c_{pq} {\mathbf e}_{p,q} \;,
\quad \mathrm{where} \quad  c_{pq} =\frac{1}{N^2} \widetilde{\varepsilon^{pl}_{pq}}
\end{equation}
In order to follow the evolution of the different modes we now
rewrite in Fourier space the argument of the $\Pos(\cdot)$ function in
the equation of evolution~(\ref{eq-mesoplast}):
 \be\ba{l} {\mathcal F}\left[
  \sigma^{ext} + G^{el}*\varepsilon^{pl}(\mathbf{r},t)
  -\sigma^{c}(\mathbf{r},\varepsilon^{pl}) \right]=
\vspace{5pt}
\\
    \sigma^{ext}\delta_p\delta_q
    +\widetilde{G^Q_{pq}} \widetilde{\varepsilon^{pl}_{pq}}
-\widetilde{\left[\sigma^c(\mathbf{r},\varepsilon^{pl}(\mathbf{r}))\right]_{pq}}
    \;.
    \ea
\ee
Ignoring for the moment the effect of the function $\Pos(\cdot)$ in Eq.~(\ref{eq-mesoplast}) we thus get by Fourier transform the evolution of the contribution of the different modes:
\be
\frac{\partial c_{pq}}{\partial t} =  \delta_p\delta_q \sigma^{ext} +\lambda_{pq} \partial c_{pq}
-\widetilde{\left[\sigma^c(\mathbf{r},\varepsilon^{pl}(\mathbf{r}))\right]_{pq}} 
\label{mode-evolution}
\ee 
This rewriting thus enables us a better understanding of the
diffusive-like behavior observed at long times for the plastic
strain. In real space, the spatial coupling is induced by the
non-local elastic interaction kernel, $G^{el}$, while the noise term
is local.  In the space of eigenmodes, the opposite character is
observed, namely the restoring force is local, but noise is not.
Since all eigenvalues are null or negatives (otherwise the dynamics
would be unstable) a competition emerges between the relaxation of the
eigenmodes induced by the elastic contribution and a random forcing
due to the quenched noise contribution.  In particular, at long times,
the contribution of the soft modes becomes dominant since they are not
submitted to relaxation. The diffusive-like behavior thus directly
emerges from a competition between the different soft modes controlled
by the quenched disorder.

The strong effect of a small MF contribution to the quadrupolar
propagator can now be re-read as the consequence of the opening of a
gap in the spectrum of eigenvalues, in other words to the vanishing of
the soft modes. In Fig.~\ref{spectrum}, the spectra of eigenvalues of
the stress redistribution kernel show the gradual gap opening due to
the introduction of a MF contribution to the elastic quadrupolar
interaction. The associated restoring elastic force brings back the
model to the standard depinning phenomenology.

Note that this interpretation only holds if we ignore the
$\Pos(\cdot)$ function that intervenes in Eq.~\ref{eq-mesoplast}.
When a long integration time in considered, the loading contributes to
a positive average that allows for such an interpretation. However, at
short time scales, the positive part function unfortunately cannot be
simply expressed in Fourier space.  A similar situation
appears in classical depinning models.  The point is that for the
latter ones, a long time integration gives a finite restoring force to
any wavelength of manifold fluctuations.  

\subsection{From Soft modes to shear-bands}

In the present context of amorphous plasticity an appealing
alternative representation of the soft modes is given by the unit
shear-bands orientated along $\pm \pi/4$. One defines ${\mathbf
  d}_{k}$ such that ${\mathbf d}_{k}(m,n)=\delta_{m-n-k}$ and
${\mathbf d'}_{k}(m,n)=\delta_{m+n-k}$ where $k \in [-N/2+1,N/2-1]$
and $\delta_n$ is the Kronecker symbol. Plastic shear bands thus
directly appear as soft modes of the quadrupolar elastic interaction,
because of the null eigenvalue, they don't induce any internal stress.

We use this decomposition to rewrite the plastic strain field  as:
\begin{equation}
    \varepsilon^{pl} = \sum_{|p| \ne |q|} c_{pq} {\mathbf e}_{p,q}
    + \sum_{k} c_k {\mathbf d}_{k} + \sum_{k} c'_k {\mathbf d'}_{k} \;,
\end{equation}
where the first sum gathers all modes of non-zero eigenvalues whereas
the two last sums correspond to combinations of shear bands oriented
at $\pm \pi/4$. Note however that the two systems of shear bands are
not independent since the scalar products ${\mathbf d}_{k}.{\mathbf
  d'}_{l}$ may be non zero. Here the amplitudes $c_k$ and $c'_k$ roughly correspond to the mean plastic strain along the shear-bands  ${\mathbf d}_{k}$ and ${\mathbf d'}_{k}$ respectively. In the same spirit as above, accounting for the non-orthogobality between the two slip systems, it is possible to write the equation of evolution of the amplitudes of the shear-bands:
 \bea
\frac{\partial c_{k}}{\partial t} +\frac{1}{N}\!\sum_\ell f_{k\ell} \frac{\partial c'_{k}}{\partial t} &= &\sigma^{ext}\!- 
\frac{1}{N}\!\sum_{\mathbf{r} \in{\mathbf d}_k } \sigma^c[(\mathbf{r},\varepsilon^{pl}(\mathbf{r})] \\
\nonumber
\frac{\partial c'_{k}}{\partial t} +\frac{1}{N}\!\sum_\ell f_{k\ell} \frac{\partial c_{k}}{\partial t} &= &\sigma^{ext}\!- 
\frac{1}{N}\!\sum_{\mathbf{r} \in{\mathbf d'}_k } \sigma^c[(\mathbf{r},\varepsilon^{pl}(\mathbf{r})]
\label{SB-evolution}
\eea
where in the present case of bands at $\pm \pi/4$, $f_{k\ell}/2 = 
  (k-l) \pmod 2$. 

As already discussed above, in absence of elastic restoring force in
the equation of evolution, we expect the strain field to be
asymptotically dominated by the sole superimposition of soft modes,
which we interpret here as shear bands at $\pm \pi/4$.

Here we obtain for the dynamics of the bands an
advection contribution due to the differnce $(\sigma^{ext}
-\overline{\sigma^{c}})$ between the driving force and the spatial
average of the threshold on the whole lattice. In addition, the
average along the bands of the fluctuating part of the thresholds and
the inter-bands coupling introduce randomness and lead to diffusion.

Note  that another important souce of interactions betwen bands
has been neglected here. Although shear bands are expected to be
dominant at long times, the short time synamics remains {\it local}. A
natural consequence of the interplay between a local threshold
dynamics and the non-local effects of the elastic interaction is the
persistence of fluctuations along the bands. The convolution of the
latter with the elastic kernel is responsible for a mechanical noise
contribution in the
dynamics~\cite{Nicolas-EPL14,Jagla-PRE15,Agoritsas-EPJE15}.


\section{Fluctuations and age statistics along shear-bands}

The interpretation of the plastic shear-bands as soft modes of the
elastic interaction encourages us to re-examine our results from this
new perspective. In particular, we expect that at long times, plastic
activity concentrates along weakly interacting shear bands. A natural
question thus arises about the respective contribution of intra
shear-bands and inter shear-bands fluctuations to the diffusive
regime. This question is reminiscent of earlier studies showing
anisotropic correlations in the plastic strain
field~\cite{Maloney-PRL09,TPVR-Meso12}.  In the same spirit, we
suggested that the long sub-diffusive regime observed in the numerical
results reflects an aging-like behavior. This motivates us to
characterize age statistics inside and outside shear-bands.

We first define the mean variance of the plastic strain field inside
the shear bands as:
\bea
W_Q^+ &= &\langle \frac{1}{N} \sum_{k=1}^{N} W_k \rangle \;, \quad \mathrm{where}\\
\nonumber
W_k &= &\frac{1}{N}
\sum_{{\mathbf r} \in {\mathbf d}_k} \left[\varepsilon^{pl} ({\mathbf r})\right]^2 
- \left[\frac{1}{N} \sum_{{\mathbf r} \in {\mathbf d}_k} \varepsilon^{pl} ({\mathbf r})
\right]^2\;.
\label{variance-SB}
\eea

In the quadrupolar geometry associated to plane shear plasticity, the
shear bands ${\mathbf d}_k$ and ${\mathbf d'}_k$ are oriented along
the $\pm \pi/4$ directions and receive a positive stress contribution
whenever one of their site experiences plasticity, hence the
superscript $+$ in the notation of the variance $W_Q^+$. In a similar
spirit we can can characterize the fluctuations of the plastic strain
field along the directions at angles $0$ and $\pi/2$ that receive a
negative stress contribution when one of their site experiences
plasticity. We denote $W_Q^-$ the variance of the plastic strain field
along such negative stress directions.

\begin{figure}
\includegraphics[width=0.95\columnwidth]{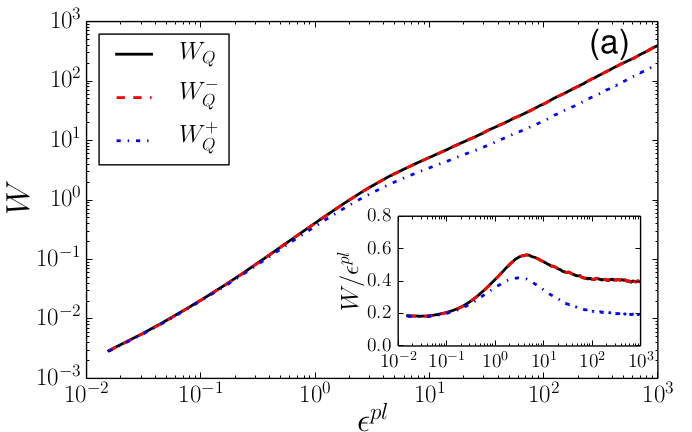}
\includegraphics[width=0.95\columnwidth]{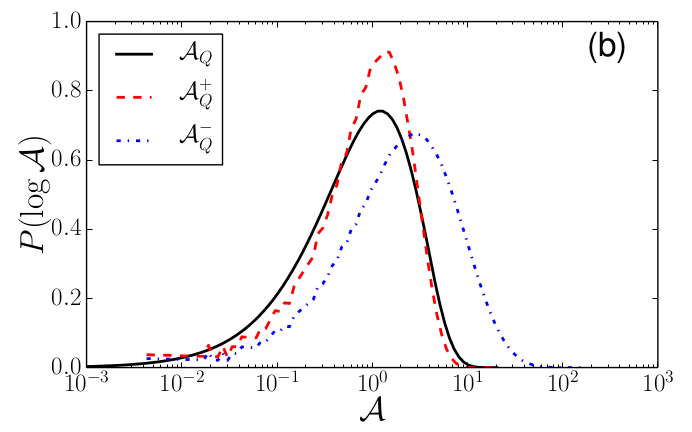}
\caption{Quadripolar kernel. (a): variance of the plastic strain field
  inside and outside shear bands. (b): Age distribution inside and
  outside shear bands}
\label{quad-bands}
\end{figure}

We show in Fig.~\ref{quad-bands}a the evolution of the global variance
$W_Q$ of the plastic strain field as well as the variances $W_Q^+$
inside the shear-bands and $W_Q^-$ outside the shear-bands. We observe
that the variance $W_Q^+$ of intra-shear-bands fluctuations are
significantly lower than the global variance $W_Q$ in the diffusion
regime. Conversely, the variance $W_Q^-$ measured in the negative
stress directions is indistinguishable from the global variance. The
inset shows the same data after rescaling by the mean plastic strain
i.e. the effective diffusivities
$D_Q=W_Q/\overline{\varepsilon_{pl}}$,
$D_Q^+=W_Q^+/\overline{\varepsilon_{pl}}$ and
$D_Q^-=W_Q^-/\overline{\varepsilon_{pl}}$. Here we see that the effective
diffusivity within the shear-bands $D_Q^+$ is about two times smaller
than the global diffusivity $D_Q$.

Beyond the spatial fluctuations, we can also characterize the temporal
fluctuations. In order to do so, we define the local age variable
${\cal A}_Q= n_A/N^2$ that counts the number of plastic events $n_A$
that occurred in the system since the last time the site has
experienced plasticity. In case of an homogeneous deformation, all
$N^2$ sites would be expected to experience plastic events at the same
frequency, hence the rescaling factor $1/N^2$. It is easy to extend
this definition to a shear-band: ${\cal A}_Q^+= n_{A^+}/N$. Here
$n_{A^+}$ is the number of plastic events since the last time a site
of the band has experienced plasticity and the rescaling factor stems
from the number $N$ of shear bands. The age ${\cal A}_Q^-$ of bands in
the negative stress directions is defined in the very same way.

We show in Fig.~\ref{quad-bands}b the distributions of ages $P(\log
{\cal A_Q})$, $P(\log {\cal A}_Q^+)$ and $P(\log {\cal A}_Q^-)$
measured in the diffusive regime. The age distribution of sites $P(\log
{\cal A_Q})$ peaks
around unity and shows a cut-off around ten. This suggests that on
average, the plastic activity is only moderately heterogeneous.

As for the
spatial fluctuations we observe that the age statistics of bands 
$P({\cal A}_Q^-)$ measured in negative stress directions (outside
shear-bands) is close to the global age statistics
$P({\cal A}_Q)$ measured on individual sites. In contrast,
the distribution $P({\cal A}_Q^+)$ of ages of the shear-bands
is shifted to larger values. A natural interpretation is that due to
the positive stress redistribution, plastic activity remains trapped
for longer periods within a shear-band (while the age of the other
bands keeps increasing) before jumping to another one. We note in
particular that the cut-off of the shear-band age distribution roughly
corresponds to the duration of the subdiffusive regime.

The spatio-temporal fluctuations of the plastic activity within the
shear-bands is thus clearly distinguishable from the backround. Still,
this diference is not dramatic. Although the diffusivity is decreased
and the duration of plastic activity is increased along the shear
bands, the qualitative picture remains unchanged. Shear bands can
survive 5-10 times longer than bands in the negative stress
directions but the age statistics ends up converging toward a
stationary distribution.  This is for instance at contrast with the
clear ergodicity breaking identified in Ref.~\cite{Torok-PRL00}.

\section{Plane vs Antiplane shear in amorphous plasticity\label{plane-vs-antiplane}}
A potential reason for the system to escape aging actually stems from
the quadrupolar geometry of the elastic interaction at play in the
present model. Since, after a plastic event, the elastic stress is
positive along the two directions at $\pm\pi/4$, it is possible to
trigger another plastic event in a direction at $0$ or $\pi$ with a
sequence of two successive events at $+\pi/4$ then $-\pi/4$ (or the
reverse). Such sequences thus restore some interaction between
positive and negative stress directions. 

In this  section we follow
this geometric idea by focussing on the case of antiplane shear
geometry early studied in Ref.~\cite{BVR-PRL02}.  As mentioned above,
in this antiplane geometry (defined in Fig~\ref{plane-antiplane}b), a
plastic inclusion induces a dipolar interaction:
\bea
G^D(r,\theta) &\approx & \frac{\cos 2\theta}{r^2}\;, \quad G^D(\mathbf{0})=-1 \\
\widetilde{G^D_{pq}} &= 
&-2A\frac{q^2}{p^2+q^2}\;, \quad \widetilde{G^D_{00}}=0
\eea

Here the soft modes are shear-bands oriented at $\theta =0$ and the
negative stress directions are oriented at $\theta =\pi$. In contrast
with the previous quadrupolar case, no direct cross-talk mechanism is
possible between the different shear-bands.
This means in particular that if we now rewrite the plastic strain field as:
\begin{equation}
    \varepsilon^{pl} = \sum_{|p| \ne 0} c_{pq} {\mathbf e}_{p,q}
    + \sum_{k} c^D_k {\mathbf d^D}_{k} \;,
\end{equation}
where the $N$ horizontal bands ${\mathbf d^D}_{k}$ are the soft modes
of the dipolar kernel $G^D$, we now obtain for the equation of
evolution of the band amplitudes:
 \be
\frac{\partial c^D_{k}}{\partial t} = \sigma^{ext} - 
\frac{1}{N} \sum_{\mathbf{r} \in{\mathbf d^D}_k } \sigma^c[(\mathbf{r},\varepsilon^{pl}(\mathbf{r})]
\ee
We thus get in the long term dynamics a set of of bands that can grow
independently of each other. Again, this statement has to be softened
to account for the effective noise induced by the sort term local
thershold dynamics that restore weak coupling between the bands.

\begin{figure}
\includegraphics[width=0.95\columnwidth]{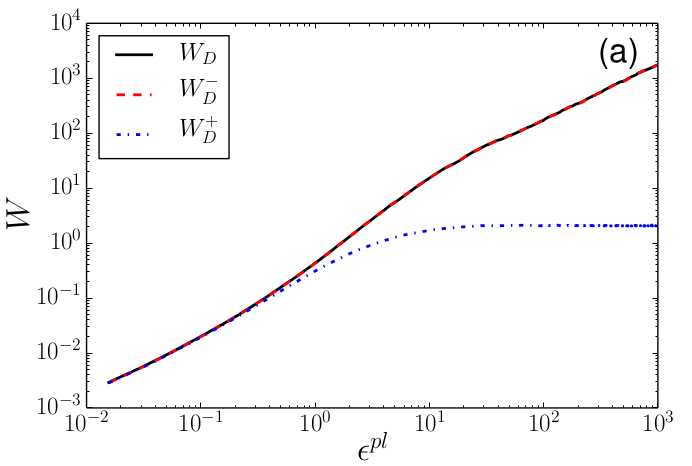}
\includegraphics[width=0.95\columnwidth]{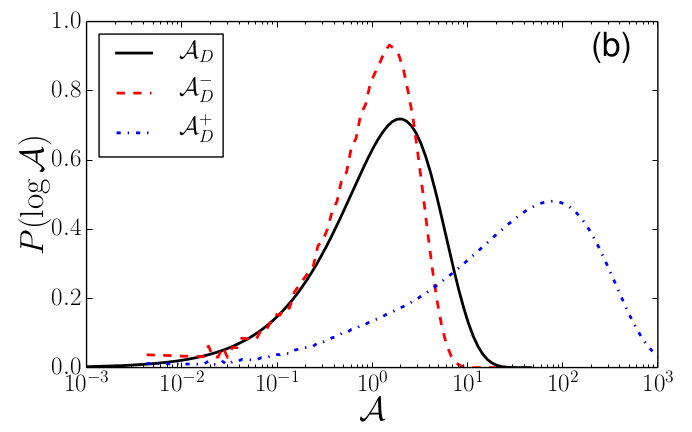}
\caption{Aging and diffusive bhaviors obtained with a Dipolar
  kernel. (a): variance $W_D$ of the plastic strain field and
  variances $W_D^+$ inside and $W_D^-$ outside shear bands. (b):
  Age distributions $P(\log {\cal A}_D)$ of the sites and and age
  distributions $P(\log {\cal A}_D^+)$, $P(\log {\cal A}_D^-)$ of the
  bands in the positive (shear-bands) and negative stress directions,
  respectively.  }
\label{dip-bands}
\end{figure}

In analogy with the previous section we show in Fig.~\ref{dip-bands}a
the evolution upon deformation of the variances $W_D^+$ and $W_D^-$ of
the plastic strain field obtained along the positive and negative
stress directions, respectively, in comparison with the global variance
$W_D$. As in the quadrupolar case, the variance $W_D^-$ in the
negative stress directions is almost the same as the global variance
$W_D$. The result is strikingly different in the direction of
shear-bands. After the power-law transient, instead of a diffusive
regime, the variance $W_D^+$ shows indeed a clear saturation. Along
the direction of the shear-bands, we thus recover the classical
Family-Vicsek phenomenology of depinning. Note however that saturation
is reached at a much later stage $\varepsilon^{pl}\approx 10$ than in
the reference Mean-Field case $\varepsilon^{pl}\approx 0.5$ (see
Fig.~\ref{diffusion1}). If one refers to the results obtained with the
composite kernels $G_a$ (see Fig.~\ref{maps-mix}), this would
correspond to small Mean-Field weight $a\approx0.005$.

In Fig.~\ref{dip-bands}b we show the distribution of ages in the
antiplane shear geometry. Again, the age distribution 
$(\log {\cal A}_D^-)$ of bands in the negative stress direction is very close to
the age distribution  $P(\log{\cal A}_D)$ of the individual
sites. The case of the shear-bands is strikingly different. Here the
age distribution $P(\log {\cal A}_D^+)$ is much older (about
two orders of magnitude) than the two other ones. One recovers the
same aging-like effect as for the shear-bands in the quadrupolar case
but with a much higher amplitude. 


\section{Conclusion}

Depinning models rely on the interplay between disorder and
  elasticity. While the yielding transition may be discussed within
  the framework of depinning, it appears that some specific properties
  of amorphous plasticity (diffusion, shear-banding) are controlled by
  the peculiar form of the quadrupolar elastic interaction. In order
  such features to be recovered in the framework of discrete lattice
  models, 
  the discretized implementation of the Eshelby kernel has to preserve
  a key property of continuum plasticity: a unit plastic strain along
  any band in a direction of maximum shear stress (here $\pm \pi/4$)
  induces no residual stress.

The interpretation of the shear-bands as soft modes of the Eshelby
elastic interaction may clarify the long debate about the relative
importance of localized rearrangements and large scale shear-band like
events~\cite{Tanguy-EPJE06},\cite{Dasgupta-PRL12b,*Ashwin-PRE13,*Dasgupta-PRE13a}
as microscopic mechanisms of amorphous plasticity and complex
rheology. It appears in particular that localized plastic events is
the rule at short time scales but on a larger time horizon, mostly
shear bands account for the kinematics, as these are the only
displacement fields that prevent large shear stress build-up.

While the present study has been concerned with the modeling of
amorphous media plasticity, a similar phenomenology is expected for
any depinning model as soon as the elastic propagator exhibits soft
modes. As discussed in~\cite{Wyart-PNAS14} in the case of plastic
yielding, this new sub-class of depinning model is expected to exhibit
non-trivial scaling properties.
More generally it is tempting to study in more details the ergodic
behavior of such models at finite temperature in relation with the
Soft Glassy Rheology models~\cite{Sollich-PRL97,*Sollich-PRE98} and
with the recent observation of the strong effect of Eshelby events on
relaxation processes in the liquid state~\cite{Lemaitre-PRL14}.

\begin{acknowledgments}
BT, SP and DV wish to express their sincere thanks to M.L. Falk and
C.E. Maloney for several stimulating discussions.
\end{acknowledgments}


%

\end{document}